\documentclass[twocolumn,showpacs,preprintnumbers,amsmath,amssymb]{revtex4}

\usepackage{graphicx}
\usepackage{amsmath}
\usepackage{bm}
\usepackage{multirow}
\usepackage{comment}

\newcommand{\simgt}{\lower.5ex\hbox{$\; \buildrel > \over \sim \;$}}
\newcommand{\simlt}{\lower.5ex\hbox{$\; \buildrel < \over \sim \;$}}
\newcommand{\largesmall}{\lower.5ex\hbox{$\; \buildrel {\LARGE >} \over {\small 
<}$}}

\begin{document}
\title{
Constraint on the growth factor of the cosmic structure from the damping
of the baryon acoustic oscillation signature}


\author{Gen Nakamura$^1$}
\email[Email:]{gen@theo.phys.sci.hiroshima-u.ac.jp}
\author{Gert H\"{u}tsi$^{2,3}$}
\email[Email:]{ghutsi@star.ucl.ac.uk} 
\author{Takahiro Sato$^1$}
\email[Email:]{sato@theo.phys.sci.hiroshima-u.ac.jp}
\author{Kazuhiro {Yamamoto}$^1$}
\email[Email:]{kazuhiro@hiroshima-u.ac.jp}

\affiliation{
$^{1}$Department of Physical Science, Hiroshima University,
Higashi-Hiroshima 739-8526,~Japan
\\
$^{2}$Department of Physics and Astronomy, University College London,
London, WC1E 6BT, UK
\\
$^{3}$Tartu Observatory, EE-61602 T\~{o}revere, Estonia}



\begin{abstract}
We determine a constraint on the growth factor by measuring the 
damping of the baryon acoustic oscillations in the matter power 
spectrum using
the Sloan Digital Sky Survey luminous red galaxy sample.
The damping of the BAO is detected at the one sigma level.
We obtain $\sigma_8D_1(z=0.3) = 0.42^{+0.34}_{-0.28}$ at the $1\sigma$ 
statistical level, where $\sigma_8$ is the root mean square overdensity
in a sphere of radius $8h^{-1}$Mpc and $D_1(z)$ is the growth factor 
at redshift $z$.  
The above result assumes that other parameters are fixed and the 
cosmology is taken to be a spatially flat cold dark matter universe 
with the cosmological constant.
\end{abstract}
\maketitle

\section{Introduction}
The baryon acoustic oscillations (BAO)
are the sound oscillations of the primeval baryon-photon fluid
prior to the recombination epoch. 
The BAO signature imprinted in the matter power spectrum  
is very useful for the study of the dark energy, hypothetically 
introduced to explain the accelerated expansion of the universe
\cite{Eis,Meik}, 
because the characteristic scale of the BAO plays a role of
a standard ruler in the universe \cite{Eis2,Seo,Matsubara}.
The BAO signature in the galaxy clustering is clearly 
detected \cite{Gert1,Percivalb}, and the constraints on the 
equation of state parameter of the 
dark energy are demonstrated \cite{Percival,Gert,Okumura,Cabre}.
Future large surveys for the precise measurement of the BAO
are in progress or planned, providing us an important tool 
to explore the origin of the accelerated expansion of the universe.

Besides the measurement of the BAO, those future surveys
provide us other important information. For example, 
the redshift-space distortions will be measured precisely 
at the same time. The redshift-space distortions reflect 
the velocity of galaxies in the direction of the line of 
sight. Especially, the linear redshift-space distortion
due to the Kaiser effect comes from the linear velocity 
field of matter perturbations, whose measurement will 
give us a chance to test the gravity theory on the cosmological 
scales \cite{Linder,Guzzo,Yamamoto,WhitePercival}. This is important
because several modified gravity models are proposed 
to explain the accelerated expansion of the universe 
as an alternative to the dark energy. Thus the measurement 
of the redshift-space distortion, which is useful to constrain the growth rate 
and the growth factor of the cosmic structure, is also important.

In Ref.~\cite{Nomura,Nomura2}, it is pointed out that the 
precise measurement of the BAO 
may also be useful to constrain the growth factor.
The BAO signature observed in the galaxy power spectrum 
is contaminated by the nonlinear effects of the density perturbations,
redshift-space distortions and the clustering bias. 
The effects of the nonlinearity and the redshift-space distortions
cause the damping of the BAO signature compared with the case
when these effects are switched off. The authors of Ref.~\cite{Nomura}
found that the damping of the BAO depends on the amplitude of the
matter power spectrum and that a measurement of the BAO damping
might be useful to constrain the growth factor of the cosmic 
structure. 

In Ref.~\cite{Nomura2}, two of the authors of the present paper
reported a detection of the BAO damping using the Sloan digital
sky survey (SDSS) luminous red galaxy (LRG) sample of the data 
release (DR) 6. Since the observed galaxy power spectrum is contaminated 
by errors, and is noisy, the procedure of extracting the BAO
signature from a galaxy power spectrum is a subtle problem.
In the previous paper~\cite{Nomura2}, the cubic spline method was 
used to construct a smooth power spectrum to extract the BAO 
signature. However, this method is very sensitive to parameters
of the cubic spline, i.e., number and interval of the nodes. 
Due to those reasons, it is clear that an independent cross check
of the previous result is useful.

In the present paper, we adopt an independent method to construct 
the smooth power spectrum, which is stable compared with the 
cubic spline method, and re-investigate the BAO damping by 
confronting the model predictions with the SDSS LRG power spectrum of the DR7.
We confirm the detection of the BAO damping at the one sigma 
level. Also we obtain a constraint on the parameter combination 
$D_1(z)\sigma_8$ at the redshift $z=0.3$ from the BAO damping
for the first time. Here $D_1(z)$ is the growth factor and 
$\sigma_8$ is the amplitude of the root mean square overdensity
in a sphere of radius $8h^{-1}$Mpc.
We denote the growth rate by $f(z)=d\ln D_1(z)/d\ln a(z)$, 
where $a(z)$ is the scale factor. 
Throughout this paper, we use units in which the velocity of light
equals 1 and the Hubble parameter $H_0=100h$km/s/Mpc with $h=0.7$.

\section{BAO damping}

The BAO signature in the power spectrum $P(k)$ can be obtained by
\begin{eqnarray}
 B(k)\equiv \frac{P(k)}{\widetilde{P}(k)}-1,
\end{eqnarray}
where $\widetilde P(k)$ represents the 'smooth' power 
spectrum corresponding to $P(k)$. 
Since the definition of
$\widetilde P(k)$ is not unique, the construction
of $\widetilde P(k)$ should be carefully done.  
The cubic spline method is frequently used \cite{Percival,Nomura2}, 
but it is not always stable if dealing with a noisy power spectrum from 
observations. 

In the present paper, we construct the smooth 
power spectrum as the linear matter power spectrum 
multiplied by a simple  analytic function of the wave number.
With the use of the constructed smooth power spectrum, we 
obtain the BAO signature from the SDSS LRG power spectrum 
of the DR7 \cite{SDSS,SDSSDR7}. The extracted BAO signature is then
compared with theoretical predictions, particularly focusing on the 
BAO damping. 

\begin{figure}[htbp]
 \includegraphics[width = 60mm,keepaspectratio,clip,angle=270]{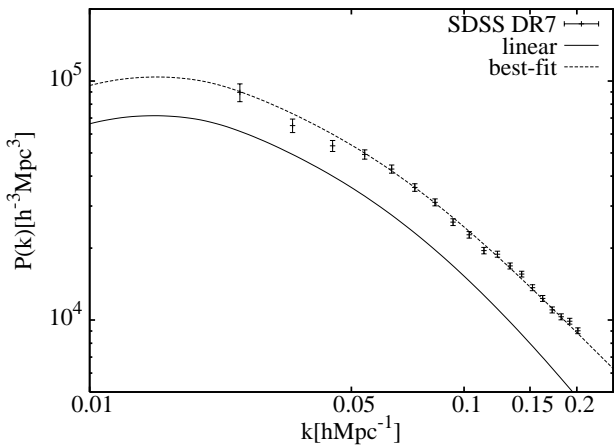}
 \caption{Comparison of the SDSS
 LRG power spectrum and the theoretical power spectra. 
 The solid curve is the linear theory  $\widetilde{P}_{\rm{lin}}(k)$.
 The dashed curve is a 'smooth' power spectrum 
 $\widetilde P_{\rm obs}(k)$ obtained with (\ref{bias}) with the Case 3.
 The cosmological parameters are described in Table \ref{tb:TAB} 
 labeled as model no.1.}
\label{fig-pow}
\end{figure}

We use the SDSS LRG sample from DR 7
(see also \cite{Percival2009,Reid2009} for recent results 
on LRGs from the SDSS DR7). Our LRG sample is restricted to the redshift range 
$z=0.16-0.47$. In order to reduce the sidelobes of the survey window we remove
some noncontiguous parts of the sample (e.g. three southern slices), which
leads us to $\sim 7150$ deg$^2$ sky coverage with a total of $100157$ LRGs.
The data reduction procedure is the same as that described in 
Ref.~\cite{Gert1}. In this power spectrum analysis, we adopted
the spatially flat Lambda cold dark matter ($\Lambda$CDM) 
model distance-redshift relation $s=s[z]$. 
Figure \ref{fig-pow} shows the observed power spectrum $P_{\rm obs}(k)$.

\begin{figure}
  \includegraphics[width = 60mm,keepaspectratio,clip,angle=270]{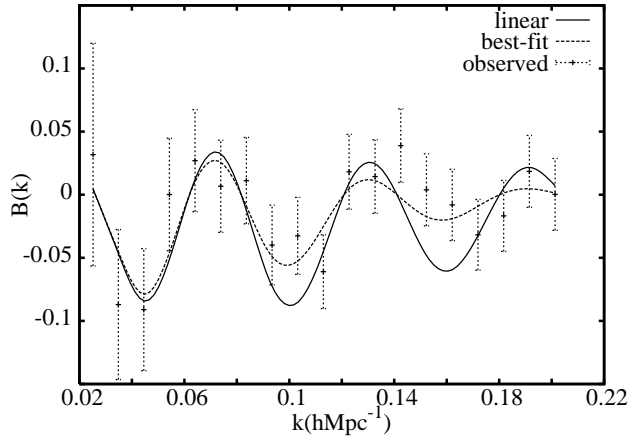}
  \caption{The BAO signature $B(k)$.
  The solid curve is the BAO in linear theory 
$B_{\rm lin}(k)$. The dashed curve is 
  the best fit theoretical curve including the damping
$B_{\rm lin}(k)(1-W(k))$. 
  The points with error bars are $B_{\rm{obs}}(k_i)$,
   the observational result obtained with the Case 3 for $A(k)$. 
  We adopted the cosmological density parameters and the
  spectral index descried in Table\ref{tb:TAB} as model no.1.}
  \label{fig-BAO}
\end{figure}

The observed power spectrum is contaminated by various effects: 
the nonlinear effect, redshift-space distortions, clustering bias,
and so on. In particular, the modeling of the clustering bias
seems to be very difficult, because it depends on the galaxy formation 
process. In the present paper, in order to construct the 
smooth power spectrum corresponding to the observed power spectrum, 
we assume 
\begin{eqnarray}
\widetilde P_{\rm obs}(k) =A^2(k)\widetilde P_{\rm lin}(k),
\label{defpobs}
\end{eqnarray}
where $\widetilde P_{\rm lin}(k)$ is the linear no-wiggle power spectrum
of \cite{Eis}.
The solid curve and the dashed curve in Fig.~\ref{fig-pow}
exemplify $\widetilde{P}_{\rm{lin}}(k)$ and $\widetilde P_{\rm obs}(k)$, 
respectively.

%
We determined the function $A(k)$ as follows: 
We first define the chi-square by
\begin{eqnarray}
\chi^2 &=&
  \sum_{i,j}\left[B_{\rm lin}(k_i)(1-W(k_i))-B_{\rm obs}(k_i)
  \right]\nonumber\\
 &&~~\times\tilde{P}_{\rm obs}(k_i){\rm
  Cov}^{-1}(k_i,k_j)\tilde{P}_{\rm obs}(k_j)\nonumber\\
 &&~~\times\left[{B_{\rm lin}(k_j)}(1-W(k_j))-
  B_{\rm obs}(k_j)\right]
 \label{bias}
\end{eqnarray}
with 
\begin{eqnarray}
&&B_{\rm lin}(k)=\frac{P_{\rm{lin}}(k)}{\widetilde{P}_{\rm{lin}}(k)}-1,
\label{defblin}
\\
&&B_{\rm obs}(k)=\frac{P_{\rm{obs}}(k)}{\widetilde P_{\rm   obs}(k)}-1,
\end{eqnarray}
where ${\rm Cov}^{-1}(k_i.k_j)$ is 
the inverse of the power spectrum covariance 
matrix \cite{Gert}, and $W(k)$ is the damping function 
(see next section for details).
In the determination of $A(k)$, we consider the following 3 cases,
\begin{description}
 \item[Case 1] We assume $A(k)=a+bk+ck^2$ and determine a, b and c 
	    by minimising (\ref{bias}) with $f=0$, $\sigma_8D_1=0$
	    (corresponding to no-damping model).
 \item[Case 2] Same as Case 1 but with $f=0.64$, $\sigma_8D_1=0.53$
	    (corresponding to the $\Lambda$CDM model with $\Omega_0=0.28$).
 \item[Case 3] We assume $A(k)=a+bk^c$ and determine a, b and c 
	    by minimizing (\ref{bias}) separately for each of the adopted cosmologies.
\end{description}

Fig.\ref{fig-BAO} shows an example of the observed BAO signature and
theoretical curves, which will be explained in the next section.
Here the cosmological parameters are the same as those of Fig.~\ref{fig-pow}.

\begin{figure}
 \includegraphics[width = 60mm,keepaspectratio,clip,angle=270]{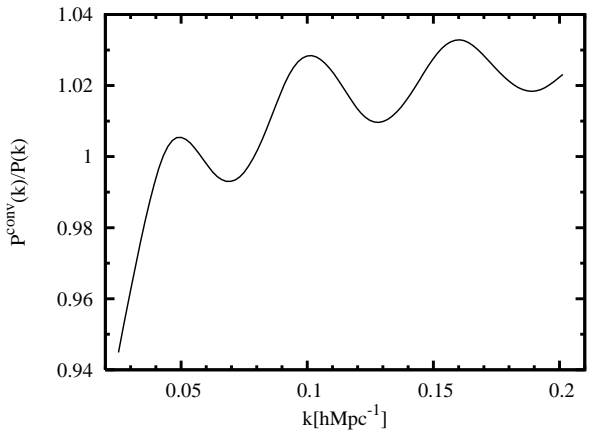}
 \caption{Ratio of the power spectrum 
convolved with the window function $P^{\rm conv}(k)$ to that without it $P(k)$.
The cosmological parameters are the same as those of Fig.\ref{fig-pow}.
}
 \label{window}
\end{figure}

Here we note the effect of the window function, which we take into 
account in our investigation. This is the finite volume effect of 
the region where galaxies are distributed, depending on the data 
analysis as well.
The effect of the window function affects the shape of the
BAO signature, which is not negligible in our investigation.
The details are described in the literature \cite{Gert1,sato}.
Fig.\ref{window} shows the ratio of the power spectrum 
convolved with the window function to that without it. 

\section{Theoretical modeling of the BAO damping}
In our theoretical modeling, we adopt the quasi-nonlinear 
power spectrum using the technique of resuming infinite 
series of higher order perturbations on the basis of the 
Lagrangian perturbation theory (LPT), which was proposed by Matsubara 
\cite{Matsubara2}. One of the advantages of using this LPT 
formalism is that the redshift-space distortions can be
taken into account. Furthermore, the predicted BAO signature
was compared with the results of N-body simulations in ~\cite{Nomura2}, 
where a good agreement was obtained. 

We denote the matter power spectrum in the redshift-space by $P(k,\mu)$,
where $\mu$ is the cosine of the angle between the line of sight 
direction and the wave number vector. 
However, the observed power spectrum in the previous section is 
the angular averaged power spectrum, which is related to $P(k,\mu)$ via
\begin{eqnarray}
 P(k)={1\over 2}\int_{-1}^{+1}P(k,\mu)d\mu.
\end{eqnarray}
Then, the BAO signature is written as
\begin{eqnarray}
 B(k)\equiv \frac{\int_{-1}^{+1}P(k,\mu)d\mu }
{\int_{-1}^{+1} \widetilde{P}(k,\mu)d\mu}-1.
\end{eqnarray}
The BAO damping is described by the function $W(k)$,
which is defined by 
\begin{eqnarray}
 B(k)=(1-W(k))B_{\rm lin}(k)
\label{Bkt}
\end{eqnarray}
where $B_{\rm lin}(k)$ is the BAO in the linear theory 
defined by Eq.~(\ref{defblin}).
Note that $B_{\rm lin}(k)$ does not depend on the redshift $z$
because the growth factor in the linear power spectrum cancels. 
We may define the angular dependent damping factor $W(k,\mu)$ by
\begin{eqnarray}
 \frac{P(k,\mu)}
{\widetilde{P}(k,\mu)}-1=\left(1-W(k,\mu)\right)B_{\rm lin}(k),
\end{eqnarray}
which leads to
\begin{eqnarray}
 W(k)=\frac{\int^{1}_{-1}d\mu W(k,\mu)\tilde{P}(k,\mu)}
  {\int^{1}_{-1}d\mu\tilde{P}(k,\mu)}.
  \label{ave-corr}
\end{eqnarray}

Following the LPT framework \cite{Matsubara}, 
the matter power spectrum in the redshift-space 
is a function of $\mu$ and the redshift $z$, which 
we denote by $P_{\rm LPT}(k,\mu,z)$. 
To calculate $P_{\rm LPT}(k,\mu,z)$ one needs the linear matter
power spectrum in redshift space, which depends on 
$\sigma_8D_1(z)$, the growth factor multiplied by the 
normalization factor, and the growth rate $f(z)=d\ln D_1(z)/d\ln a(z)$, 
where $a(z)$ is the scale factor. 
The full expression for $P_{\rm LPT}(k,\mu,z)$, which is rather complicated, 
can be found in Ref.~\cite{Matsubara}. 

In the previous work \cite{Nomura2}, it was found that the 
angular dependent damping 
function at the redshift $z$ is approximately given by 
\begin{eqnarray}
{\cal W}(k,\mu;z)=\frac{D_1^2(z)}{1+\alpha(\mu,z)D_1^2(z)\tilde{g}(k)}
           \frac{\tilde{P}^{(\rm{s})}_{22}(k,\mu)}{\tilde{P}^{(\rm{s})}_{\rm{lin}}(k,\mu)},
 \label{corr}
\end{eqnarray}
where ${P}^{(\rm{s})}_{\rm{lin}}(k,\mu)=(1+f\mu^2)^2P_{\rm lin}(k)$,
$\alpha(\mu,z)=1+f(f+2)\mu^2$,
${g}(k)=(k^2/6\pi^2)\times\int_0^\infty dq P_{\rm lin}(q)$, and 
${{P}_{(\rm{s})22}(k,\mu)}$ is the function to describe the higher 
order corrections in the LPT framework, which is given in Appendix B
of Ref.~\cite{Matsubara}, but note that the 'tilde' means the quantity 
with the no-wiggle linear power spectrum. 

\begin{figure}[htbp]
\begin{center}
\includegraphics[width = 70mm,keepaspectratio,clip,angle=270]{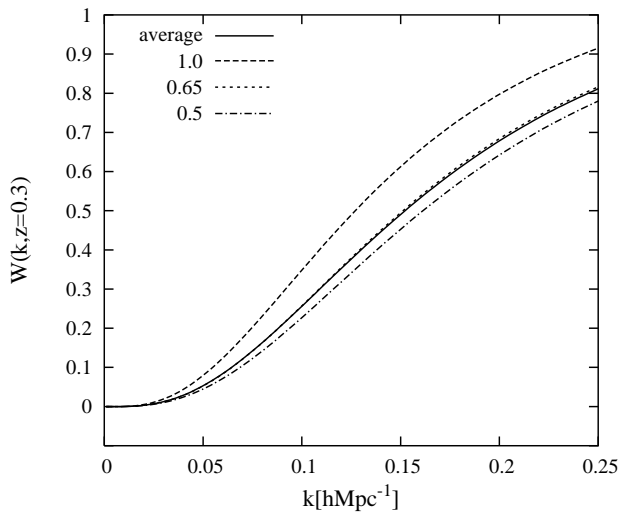}
\caption{The damping function ${\cal W}(k,\mu;z)$ at redshift z=0.3 as
a function of the wave number, for $\mu=0.5$ (dot dashed), 
$0.65$ (dotted) and $1.0$ (long dashed), respectively.
The solid curve is the angular averaged damping function Eq.~(\ref{ave-corr}).
The cosmological parameters are described in Table\ref{tb:TAB} labeled as 
model no. 1.}
\label{fig-corr}
\end{center}
\end{figure}

In the present paper, we adopt ${\cal W}(k,\mu=0.65,z=0.3)$ as the 
theoretical damping function $W(k)$ to be compared with observations. 
Here $z=0.3$ is a mean redshift of the LRG sample. Also, the angular averaged 
damping function defined by Eq.~(\ref{ave-corr}) is well
approximated by ${\cal W}(k,\mu=0.65,z=0.3)$, which is 
demonstrated in Figure \ref{fig-corr}.
The solid curve is the angular averaged damping function Eq.~(\ref{ave-corr}),
the dotted curve is  ${\cal W}(k,\mu=0.65,z=0.3)$.

In summary, the damping function $W(k)$ depends on the linear 
matter power spectrum and the growth rate $f(z)$ at the mean 
redshift $z=0.3$. Then, by comparing the BAO damping with 
observations, we can hope to constrain $\sigma_8D_1(z)$ and $f(z)$.

\begin{figure}[htbp]
\begin{center}
  \includegraphics[width = 60mm,keepaspectratio,clip,angle=270]{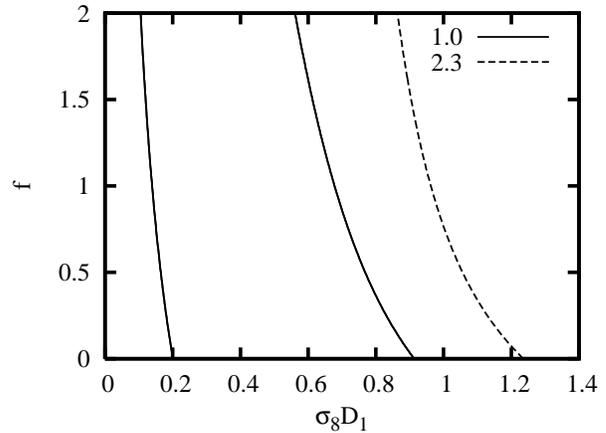}
  \caption{The contour of $\Delta\chi^2$ =1.0, 2.3 on the $\sigma_8D_1$-$f$
  plane with Case 3. The other cosmological parameters are fixed as
  those of model no. 1. in Table \ref{tb:TAB}.}
  \label{fig-contour}
\end{center}
\end{figure}


\section{Comparison between observation and theory}
We compare the observational result and the theory by 
calculating the chi-square defined by Eq.~(\ref{bias})
with the function $A(k)$ determined by the methods in section 2.
In our analysis, we treat $f$ and $\sigma_8 D_1$
as independent variables which have nothing
to do with the cosmological parameters.
%
Figure \ref{fig-contour} is the contour of $\Delta \chi^2$ 
at 1$\sigma$ level on the $\sigma_8D_1$-$f$ plane,
where the other parameters are fixed as $\Omega_0=0.28$, 
$\Omega_b=0.046$ and the spectral index $n_s=0.96$.
The best fit value is $\sigma_8D_1=0.40$ and $f=0.76$, but 
the BAO damping is not very sensitive to the growth rate $f$.

\begin{figure}[htbp]
\begin{center}
  \includegraphics[width = 60mm,keepaspectratio,clip,angle=270]{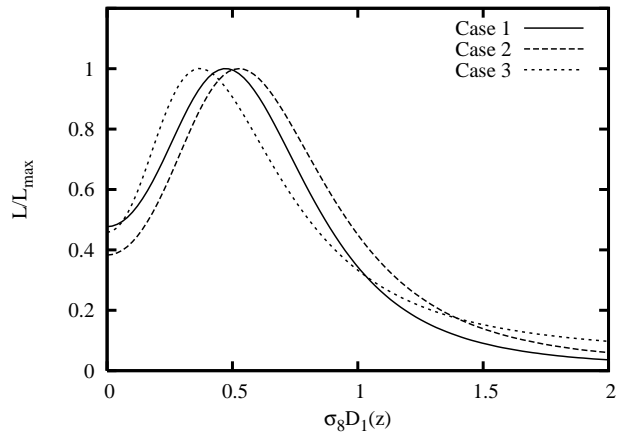}
 \caption{The likelihood as a function of $\sigma_8D_1$ 
  with $f$ integrated over the range $0\leq f\leq 2$.  
  The cosmological parameters are fixed as
  those of model no. 4. in Table \ref{tb:TAB}.}
  \label{fig-like}
\end{center}
\end{figure}

Figure \ref{fig-like} shows the likelihood as a function of 
$\sigma_8D_1$, where the parameter $f$ is integrated over 
the range $0\leq f\leq 2$.

Table \ref{tb:TAB} shows the results of the chi-squared test 
Eq.~(\ref{bias}) for various cosmological parameters.
For the consistency of the analysis, we used the data
of $P_{\rm obs}(k)$, which was calculated adopting 
the distance redshift relation $s=s[z]$ of the 
same cosmological parameters.
From this table, it is clear that finite positive values of
$\sigma_8D_1>0$ better fit the data for all the models. 
This allows us to conclude that the detection of the BAO damping 
is at the 1 $\sigma$ level at the worst case.
The $\Lambda$CDM model with $\Omega_0=0.28$ predicts $D_1(z=0.3)
\sim 0.65$. If we adopt the value of $\sigma_8=0.8$ from 
Ref.~\cite{Komatsu}, we have $\sigma_8D_1\sim 0.53$.
This is consistent with our result within the 1$\sigma$ level.

To discuss the attainable constraints on $\sigma_8D_1(z)$ in future observations,
we calculate the relevant Fisher matrix component   
\begin{eqnarray}
 F = \frac{1}{4\pi^2}\int^{k_{max}}_{k_{min}}
  dk k^2\left(\frac{\partial B(k,z)}{\partial (\sigma_8D_1)}\right)^2
  \frac{\tilde{P}_{\rm gal}^2(k,z)}{{\cal Q}^2(k,z)},
\end{eqnarray}
where we set $k_{\rm max}=0.2,k_{\rm min}=0.02,
B(k,z)=(1-{\cal W}(k,\mu=0.65),z)B_{\rm lin}(k)$, 
$\tilde{P}_{\rm gal}(k,z)$ is the galaxy power spectrum 
and ${\cal Q}(k,z)=\Delta A\int_{\rm z_{\rm min}}^{\rm z_{\rm max}} 
dz (ds/dz)s^2 \bar n^2
/[1+\bar n P_{\rm gal}(k,z)]^2$ with the mean number density
of galaxies $\bar n$ and the survey area $\Delta A$.
We adopted the Q-model for the galaxy power spectrum,
elaborated in \cite{Sanchez} with the parameters $A_1=1.4$ and $Q=16$
(see also \cite{Nomura}).

\begin{figure}[htbp]
\begin{center}
  \includegraphics[width = 60mm,keepaspectratio,clip,angle=270]{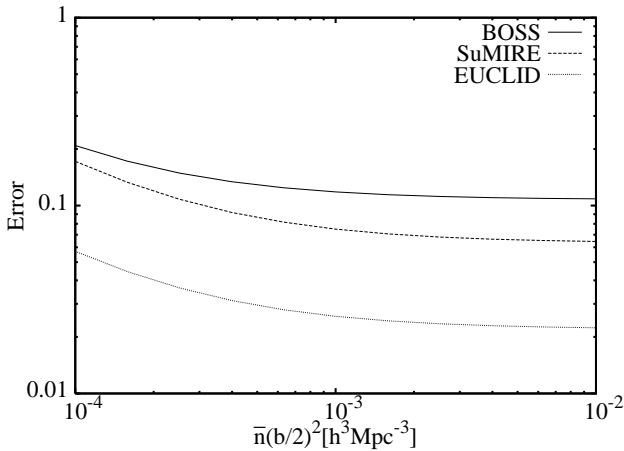}
 \caption{The 1$\sigma$-level statistical errors of $\sigma_8D_1(z)$
 as a function of $\bar{n}(b/2)^2$ with the constant number density 
$\bar n$ and the bias $b$.
 }
  \label{fig-fisher}
\end{center}
\end{figure}

The minimum error attainable on $\sigma_8D_1$ is given by $1/F^{1/2}$.
Figure \ref{fig-fisher} shows $1/F^{1/2}$ as a function of 
$\bar{n}(b/2)^2$, where $b$ is the bias parameter of the Q-model.
For the future surveys, we adopted the parameters in Table II 
\cite{WhitePercival,BOSS,SuMIRE}.
We computed the constraint at the mean redshift $z=(z_{\rm max}+z_{\rm min})/2$ 
for each survey. 
The minimum error $\sigma_8D_1$ is typically $0.1$, but depends 
on the mean number density and the clustering bias.


\section{SUMMARY AND CONCLUSIONS}
In this work, we examined the BAO damping.
We used the observed power spectrum of the SDSS LRG sample of DR7. 
The result shows that the BAO damping really exists and a
constraint on $\sigma_8D_1$ is obtained for the first time from 
the damping. 
The detection of the BAO damping is robust, and do not 
depend much on the specific model used to define the smooth model.
The constraint on $\sigma_8D_1$  is note very tight,  
but this is an independent and unique test for the growth factor.
Furthermore, this method might be useful for future surveys 
\cite{Nomura2}.

In the present analysis, we have not considered the effect of 
the clustering bias on the BAO damping. This point might be
considered more carefully in the future.

\vspace{3mm}
\section*{Acknowledgements} 
We thank H.~Nomura and T.~Nishimichi for useful suggestions
and comments.
This work was supported by a Grant-in-Aid
for Scientific research of Japanese Ministry of Education, 
Culture, Sports, Science and Technology (No.~21540270).
This work is supported in part by Japan Society for Promotion
of Science (JSPS) Core-to-Core Program, International Research 
Network for Dark Energy.
We used the CAMB code for computation of the wiggle power 
spectrum \cite{CAMB}.

\begin{table*}[b]
\begin{tabular}{ c@{\hspace{1cm}} c@{\hspace{1cm}} c@{\hspace{1cm}}
 c@{\hspace{1cm}} c@{\hspace{1cm}} c@{\hspace{1cm}}  c@{\hspace{1cm}} c}
\hline
\hline
No. &$\Omega_m$ &$\Omega_b$ &$n_s$ &f &$\sigma_8D_1$($1\sigma$
 level) Case 1&Case 2&Case 3 \\ 
\hline
1 & 0.28& 0.046 & 0.96 & 0.64 &0.38$^{+0.26}_{-0.26}$ &0.42$^{+0.30}_{-0.24}$ &0.42$^{+0.34}_{-0.28}$  \\
2 & 0.28& 0.046 & 0.96 & 0.60 &0.38$^{+0.27}_{-0.25}$ &0.42$^{+0.30}_{-0.24}$ &0.42$^{+0.33}_{-0.27}$  \\
3 & 0.28& 0.046 & 0.96 & 0.70 &0.37$^{+0.26}_{-0.24}$ &0.40$^{+0.30}_{-0.22}$ &0.41$^{+0.32}_{-0.26}$  \\
4 & 0.28& 0.046 & 0.96 & {\rm integrated}&0.48$^{+0.30}_{-0.30}$ &0.52$^{+0.36}_{-0.30}$ &0.38$^{+0.38}_{-0.28}$ \\
5 & 0.27& 0.046 & 0.96 & 0.64 &0.44$^{+0.25}_{-0.21}$ &0.46$^{+0.28}_{-0.20}$   &0.56$^{+0.40}_{-0.24}$   \\
6 & 0.29& 0.046 & 0.96 & 0.64 &0.36$^{+0.32}_{-0.32}$ &0.42$^{+0.35}_{-0.28}$&0.50$^{+0.36}_{-0.26}$   \\
7 & 0.28& 0.048 & 0.96 & 0.64 &0.38$^{+0.26}_{-0.21}$ &0.42$^{+0.28}_{-0.21}$ &0.42$^{+0.32}_{-0.22}$  \\
8 & 0.28& 0.044 & 0.96 & 0.64 &0.36$^{+0.29}_{-0.28}$ &0.42$^{+0.31}_{-0.27}$ &0.40$^{+0.36}_{-0.30}$  \\
9 & 0.28& 0.046 & 0.98 & 0.64 &0.40$^{+0.30}_{-0.24}$ &0.46$^{+0.32}_{-0.26}$ &0.42$^{+0.33}_{-0.26}$   \\
10& 0.28& 0.046 & 0.94 & 0.64 &0.34$^{+0.25}_{-0.24}$ &0.38$^{+0.27}_{-0.22}$ &0.40$^{+0.32}_{-0.26}$  \\
\hline
\hline
\end{tabular}
\caption{The results of the chi-squared test of the BAO
damping for various cosmological models. In the table, 
'integrated' means that $f$ is integrated over the range $0\leq f\leq 2$.}
\label{tb:TAB}
\end{table*}

\begin{table*}[b]
\begin{tabular}{ c@{\hspace{1cm}} c@{\hspace{1cm}} c@{\hspace{1cm}}
 c@{\hspace{1cm}} c}
\hline
\hline
 {\rm Survey} &$z_{\rm min}$ &$z_{\rm max}$
 &$\bar{n}[h^3\rm{Mpc}^{-3}]$(Tyipical value) &${\rm Area}~\Delta A [{\rm deg.}^2]$ 
 \\ 
\hline
BOSS   & $0.1$&$ 0.7 $&$3\times 10^{-4}$ & $ 10^4$ \\
SuMIRE & $0.7$&$ 1.6 $&$4\times 10^{-4}$ & $2\times 10^3$ \\
EUCLID & $0.1$&$ 2   $&$5\times 10^{-3}$ & $2\times 10^4$ \\
\hline
\hline
\end{tabular}
\caption{Future survey parameters adopted in the Fisher matrix analysis.}
\label{tb:TABII}
\end{table*}

\end{document}